\documentclass[12pt,nofootinbib]{revtex4}
\usepackage{amssymb,epsf,amsmath}

\newcommand{\oR}{{\mathbb R}}
\newcommand{\oV}{{\mathbb V}}
\newcommand{\oC}{{\mathbb C}}

\newcommand{\oG}{{\mathbb G}}

\newcommand{\bi}{{\boldsymbol{\iota}}}
\newcommand{\bt}{{\boldsymbol{\theta}}}

\renewcommand{\Im}{\mathop{\mathrm{Im}}\nolimits}

\newcommand{\spec}{\mathop{\mathrm{sp}}\nolimits}

\newcommand{\supp}{\mathop{\mathrm{supp}}\nolimits}
\newcommand{\eqdef}{\stackrel{\mathrm{def}}{=}}

\newtheorem{theorem}{Theorem}
\newtheorem{lemma}{Lemma}
\newtheorem{remark}{Remark}

\newcommand{\labelT}[1]{\label{T#1}}

\begin{document}

\title{Wedge locality and asymptotic commutativity}

\author{M.~A.~Soloviev}
\email{soloviev@lpi.ru}
\affiliation{I.~E.~Tamm Department of Theoretical Physics,
P.~N.~Lebedev Physical Institute,\\ Russian Academy of Sciences,
Leninsky Prospect~53,  Moscow 119991, Russia}

\begin{abstract}
\medskip
\baselineskip=15pt
In this paper, we study twist deformed quantum field theories
obtained by combining the Wightman axiomatic approach with the
idea of spacetime noncommutativity. We prove that the deformed
fields with deformation parameters of opposite sign satisfy the
condition of mutual asymptotic commutativity, which was used
earlier in nonlocal quantum field theory as a substitute for
relative locality.  We also present an improved proof of the wedge
localization property discovered for the deformed  fields by
Grosse and Lechner,  and we show that  the deformation leaves the
asymptotic behavior of the vacuum expectation values in spacelike
directions substantially unchanged.
\end{abstract}

\maketitle


\baselineskip=20pt

\section{Introduction}\label{S1}

Models of quantum field theory on noncommutative spacetime
continue to attract attention because of their relevance for
understanding quantum gravity~\cite{DFR} and because they can be
obtained as a particular low-energy limit of  string
theory~\cite{SeiW}. Noncommutativity is usually introduced by
replacing the spacetime coordinates $x^\mu$  with Hermitian
operators $\hat x^\mu$ satisfying  commutation relations of the
form
\begin{equation}
[x^\mu,  x^\nu] =i\theta^{\mu\nu},
 \label{1.1}
\end{equation}
where $\theta^{\mu\nu}$ is a real antisymmetric matrix, constant
in the simplest case. The relations~\eqref{1.1} are translation
invariant, but not Lorentz covariant. The twist deformation was
devised~\cite{ChKNT,W} as a way to restore the spacetime
symmetries broken by noncommutativity.  In its widest
form~\cite{ALV}, the twisting principle implies  that all
symmetries and products of the theory should be consistently
deformed by properly applying a twist operator. In particular, the
tensor product $f\otimes g$ of two functions on spacetime is
deformed  in the following way: $f\otimes g\to f\otimes_\theta g$,
where
\begin{equation}
(f\otimes_\theta g)(x,y)\eqdef\exp\left(\frac{i}{2}\theta^{\mu\nu}
\frac{\partial}{\partial x^\mu}\frac{\partial}{\partial
y^\mu}\right)f(x)g(y),
 \label{1.2}
\end{equation}
and the twist operator here is
\begin{equation}
\mathcal
T=e^{\tfrac{i}{2}\theta^{\mu\nu}\partial_\mu\otimes\partial_\nu}.
 \label{1.3}
\end{equation}
(Hereafter, we use the usual summation convention for the repeated
indices.) From the standpoint of deformation quantization,
noncommutativity amounts to deforming the ordinary pointwise
product $f(x)g(x)$ to the Weyl-Moyal star product $f\star_\theta
g$ which is obtained from $f\otimes_\theta g$ by restricting to
the diagonal,
\begin{equation}
(f\star_\theta g)(x)=(f\otimes_\theta g)(x,x).
 \label{1.4}
\end{equation}
For the coordinate functions, we have
\begin{equation}
[x^\mu,x^\nu]_\star\equiv x^\mu\star x^\nu-x^\nu\star
x^\mu=i\theta^{\mu\nu},
 \label{1.5}
\end{equation}
which is related to~\eqref{1.1} by the Weyl-Wigner correspondence.
The strategy of twisting also leads to deformed commutation
relations for the creation and  annihilation operators of free
fields, see~\cite{ALV,Z,BGMPQV,FW,RS,JM}. However, as shown
in~\cite{FW,JM,BPQ}, some combinations of twistings can cancel
noncommutativity. Then the S-matrix of twisted quantum field
theory turns out to be equivalent to that of its commutative
counterpart, and this issue does not seem completely resolved
(compare, e.g.,~\cite{F} and \cite{ChNST}).

A new interesting line of research concerns the use of
noncommutative deformations of  free field theories as a means of
constructing integrable models with a  factorizable  S-matrix.
Grosse and Lechner~\cite{GL1,GL2} studied a deformation of this
type, generated by  twisting the tensor algebra of test functions
in  the Wightman framework~\cite{SW}, and they discovered that the
deformed fields can be localized in wedge-shaped regions of
Minkowski space. Grosse and Lechner also showed that the
deformation introduces a nontrivial interaction and that this
weak form of locality is sufficient for computing two-particle
S-matrix elements. A more general deformation techniques have been
developed in an operator-algebraic setting~\cite{BLS,L11} and then
extended to quantum field theory on a curved spacetime~\cite{DLM}.
This deformation method was also applied to a fermionic
model~\cite{A} and was used to construct wedge-local fields with
anyonic statistics~\cite{P}.

In this paper, we consider the twisted  quantum field theory from
a complementary point of view stated in~\cite{S08,S10}, with
emphasis on the nonlocal aspects of the deformation. The
deformation procedure described below in terms of the Wightman
functions applies to interacting as well as free fields. Our main
observation is that the fields $\phi^\theta$ and $\phi^{-\theta}$
with deformation parameters of opposite sign satisfy the condition
of mutual asymptotic commutativity, which was used earlier in
nonlocal quantum field theory (see, e.g.,~\cite{S99} and
references therein) as an analog of relative locality. This result
supplements the wedge localization property found
in~\cite{GL1,GL2}. We also show that the deformation does not
spoil the asymptotic behavior of the  vacuum expectation values in
spacelike directions, which plays the major role in constructing
the asymptotic scattering states in the deformed theory.

The paper is organized as follows. In Sec.~\ref{S2}, we  list
basic properties of the twisted tensor product $\otimes_\theta$
and define the corresponding deformation of Wightman functions. In
Sec.~\ref{S3}, we present an improved proof of the wedge locality
property of the deformed fields. In Sec.~\ref{S4}, we obtain a
characterization of the asymptotic behavior of the
(anti)commutator
$[\phi^\theta(x),\phi^{-\theta}(y)]_{\stackrel{-}{(+)}}$ of two
fields with deformation parameters of opposite sign. We show that
this commutator falls off rapidly at large spacelike separation of
$x$ and $y$, and we estimate  the fall-off rate. Particular
attention is given to the adequate choice of the test functions
that are required for this purpose. At this point, we use  a
criterion~\cite{S07TMP} under which a test function space has the
structure of  an algebra with respect to the Weyl-Moyal star
product. Our analysis shows, in particular, that the commutator
under study satisfies the asymptotic commutativity condition
proposed for nonlocal fields in~\cite{FS}. In Sec.~\ref{S5}, we
prove that the deformation has little or no effect on the
asymptotic behavior of the   vacuum expectation values in
spacelike directions. Section~\ref{S6} contains concluding
remarks.

\section{Twist deformation of Wightman functions}\label{S2}

In order that the twisted tensor product~\eqref{1.2} and
Weyl-Moyal star product~\eqref{1.4} be well defined, the functions
involved must satisfy certain conditions. In quantum field theory
formalism, it is a standard practice to use the Schwartz space $S$
of smooth functions  decreasing faster than any inverse power of
their arguments, and this space is, as well known, an algebra
under the star multiplication. But it should be kept in mind that
the expansions of both these products in powers of the
noncommutative parameter $\theta$ are in general divergent for
functions in $S$. A preferable definition  of these products is by
using the Fourier transformation, which converts the twist
operator~\eqref{1.3} to the multiplication by the function
\begin{equation}
\eta(p,q)= e^{-\tfrac{i}{2}p\theta q}, \qquad\text{where}\quad
p\theta q\eqdef p_\mu\theta^{\mu\nu}q_\nu.
 \label{2.1}
\end{equation}
The function $\eta$ is  a multiplier of the Schwartz space and
$\widehat{f\otimes_\theta g}$ may be written as
\begin{equation}
(\widehat{f\otimes_\theta g})(p,q)= e^{-\tfrac{i}{2}p\theta q}\hat
f(p)\hat g(q).
 \label{2.2}
\end{equation}
This definition extends to the case of several variables in the
following way
\begin{multline}
(\widehat{f_{(m)}\otimes_\theta g_{(n)}})(p_1,\dots
p_m;q_1,\dots,q_n) =\\=
\prod_{j=1}^m\prod_{k=1}^ne^{-\tfrac{i}{2}p_j\theta q_k} \hat
f_{(m)}(p_1,\dots p_m) \hat g_{(n)}(q_1,\dots,q_n), \label{2.3}
\end{multline}
where $f_m$ and $g_n$ are assumed to be elements  of $S(\oR^{4m})$
and $S(\oR^{4n})$ respectively. It is easy to see that the
bilinear map $(f,g)\to f\otimes_\theta g$ is continuous in the
topology of the Schwartz space and satisfies the associativity
condition $f\otimes_\theta (g\otimes_\theta h) =(f\otimes_\theta
g)\otimes_\theta h$, which really determines  the form of the
multiplier in~\eqref{2.3}.

We now turn to the noncommutative deformation~\cite{GL1,GL2,S08}
of quantum field theories that can be associated with the twisted
tensor product. Let $\{\phi_\iota\}_{\iota\in I}$ be  a finite
system of quantum  fields  transforming according to irreducible
finite-dimensional representations of the proper Lorentz group
$L_+^\uparrow$ or its covering group $SL(2,\oC)$. Their components
are labelled by an additional index $l$, but for  brevity we let
$\bi$ denote the pair $(\iota,l)$. We suppose that all the
assumptions of the Wightman formulation~\cite{SW} of local quantum
field theory are satisfied and $\phi_\bi$ are defined as
operator-valued distributions with a common dense invariant domain
in a Hilbert space $\mathcal H$. As usual, we denote by $\Omega$
the vacuum state, by $w_{\bi_1\dots\bi_n}$ the vacuum expectation
value of a product of $n$ fields, and identify it with a tempered
distribution on $\oR^{4n}$,
\begin{equation}
\langle \Omega,
\phi_{\bi_1}(f_1)\cdots\phi_{\bi_n}(f_n)\Omega\rangle=
w_{\bi_1\dots\bi_n}(f_1\otimes\cdots\otimes f_n), \qquad
w_{\bi_1\dots\bi_n}\in S'({\mathbb R}^{4n}).
 \label{2.4}
\end{equation}
The deformed Wightman functions  $w_{(\dots)}^\theta$ are defined
by
\begin{equation}
w^\theta_{\bi_1\dots\bi_n}(f_1\otimes\cdots\otimes f_n)\eqdef
w_{\bi_1\dots\bi_n}(f_1\otimes_\theta\cdots\otimes_\theta f_n),
\qquad f_j\in S(\oR^4),
 \label{2.5}
\end{equation}
or equivalently, by
\begin{equation}
\hat w^\theta_{\bi_1\dots\bi_n}= \prod_{1\leq j<k\leq n}
e^{-\tfrac{i}{2}p_j\theta p_k}\hat w_{\bi_1\dots\bi_n}.
 \label{2.6}
\end{equation}
The set of deformed distributions  $w^\theta_{(\dots)}$ satisfies
the positive-definiteness condition (see~\cite{S08}). Furthermore,
as shown below in Sec.~\ref{S5}, the deformation~\eqref{2.5} does
not spoil the cluster decomposition property, and so if $a$ is a
spacelike vector, then for any  $f\in S(\oR^{4m})$ and $g\in
S(\oR^{4(n-m)})$, the following relation holds:
\begin{equation}
w^\theta_{\bi_1\dots\bi_n}(f\otimes g_{(\lambda
a)})\longrightarrow
w^\theta_{\bi_1\dots\bi_m}(f)w^\theta_{\bi_{m+1}\dots\bi_n}(g)
\qquad \mbox{as}\quad \lambda\to \infty,
 \label{2.7}
\end{equation}
where by $g_{(\lambda a)}$ we mean the shifted function, i.e.,
$g_{(\lambda a)}(x_{m+1},\dots, x_n)=g(x_{m+1}-\lambda
a,\dots,x_n-\lambda a)$. Therefore, by the Wightman reconstruction
theorem~\cite{SW}, this set of distributions determines  a field
theory uniquely, up to unitary equivalence. It is easy to
construct explicitly quantum fields $\phi^\theta_\iota$ having
such expectation values. As shown in~\cite{SW}, the Schwartz
kernel theorem gives a precise meaning to vectors of the form
\begin{equation}
\Phi_{\bi_1\dots\bi_n} (g)=\int dx_1\dots dx_n\,g(x_1, \dots, x_n)
\phi_{\bi_1}(x_1)\cdots\phi_{\bi_n}(x_n)\Omega,\quad\text{where
$g\in S({\mathbb R}^{4n})$}, \label{2.8}
\end{equation}
and the linear  subspace $D$ spanned by all these vectors and
$\Omega$ can be taken as a common domain of the initial  fields
$\phi_\iota$. For each $f\in S(\oR^4)$, we define
$\phi^\theta_\bi(f)$ by
\begin{equation}
\phi^\theta_\bi(f)\Omega=\phi_\bi(f)\Omega,\qquad
\phi^\theta_\bi(f)\Phi_{\bi_1\dots\bi_n}(g)=
\Phi_{\bi\bi_1\dots\bi_n}(f\otimes_\theta g),\quad n\geq
1,\label{2.9}
\end{equation}
extended by linearity.

It is easy to verify that the fields  $\phi^\theta_\bi(f)$ are
well defined as operator-valued tempered distributions with the
same common domain $D\subset\mathcal H$, and it is clear that
\begin{equation}
\langle
\Omega,\,\phi^\theta_{\bi_1}(f_1)\cdots\phi^\theta_{\bi_1}(f_n)\Omega\rangle=
w^\theta_{\bi_1\dots\bi_n}(f_1\otimes\cdots\otimes f_n). \notag
 \end{equation}
We also note that the linear span of all vectors of the form
analogous to~\eqref{2.8} but with  $\phi^\theta_{\bi_j}$ in place
of $\phi_{\bi_j}$ coincides with  $D$, because the multiplier
$\prod_{j<k} e^{-\tfrac{i}{2}p_j\theta p_k}$ maps $S({\mathbb
R}^{4n})$ isomorphically onto itself.

The basic properties of the deformed Wightman functions and fields
are described in~\cite{GL1,GL2} and in~\cite{S08,S10} for the case
of a single neutral scalar field. The deformation does not change
the  support properties of the vacuum expectation values in the
momentum space, and therefore the distributions
$w^\theta_{(\dots)}$ satisfy the spectrum condition. The
translation invariance is also preserved. The vacuum $\Omega$ is a
cyclic vector for the deformed fields $\phi^\theta_\iota$.
Moreover, as pointed out in~\cite{GL2}, they have the
Reeh-Schlieder property, i.e., for each nonempty open set
$O\subset \oR^4$, the linear span of vectors of the form
$\prod_{j=1}^n\phi^\theta_{\bi_j}(f_j)\Omega$ with $\supp
f_j\subset O$ is dense in $\mathcal H$. If a field $\phi_\bi$ is
Hermitian, then so is $\phi_\bi^\theta$. The derivation of these
properties uses in an essential way the identity
\begin{equation}
w_{(\dots)}(f\otimes_\theta g)=w_{(\dots)}(f\otimes g),
\label{2.10}
\end{equation}
which holds for every   $n$-point Wightman function and for any
test functions  $f\in S(\oR^{4m})$ and $g\in S(\oR^{4(n-m)})$,
where $1<m<n$. The identity~\eqref{2.10} follows directly
from~\eqref{2.3} and the translation invariance of the
distributions $w_{(\dots)}$, because the matrix  $\theta^{\mu\nu}$
is antisymmetric. In the case of a free  neutral scalar  field
$\phi$,  its creation and annihilation operators are deformed as
follows
\begin{equation*}
a_\theta(p)=e^{\tfrac{i}{2}p\,\theta P} a(p),\qquad
a^*_\theta(p)=e^{-\tfrac{i}{2}p\,\theta P} a^*(p),
\end{equation*}
where  $P$ is the  energy-momentum operator. The operators
$a_\theta(p)$ and  $a^*_\theta(p)$ satisfy the deformed canonical
commutation relations discussed in~\cite{ALV,Z,BGMPQV,FW,RS,JM}. As
already noted, the deformation~\eqref{2.5} preserves the
translation invariance, but it violates the Lorentz covariance and
the fields $\phi_\iota^\theta$ transform covariantly only under
those Lorentz trans\-formations that leave the matrix
$\theta^{\mu\nu}$ unaltered. This deformation also leads to a
strong violation of locality, and the fields $\phi_\iota^\theta$
do not satisfy the microcausality condition. This is easily seen
by considering matrix elements of the deformed field commutator in
the simplest case of a free scalar field.  Theorem~3 of~\cite{S08}
shows that if $\theta^{\mu\nu}\ne0$, then the matrix elements of
the form $\langle \Omega, [\phi^\theta(x),\phi^\theta(y)]\,
\Phi\rangle$, where $\Phi$  is a normalized two-particle state,
are  nonzero everywhere, i.e., their supports coincide with
$\oR^4\times\oR^4$. It should be noted that this is also true for
the case of so-called space-space noncommutativity, where
$\theta^{0\nu}=0$ for all $\nu$. Therefore the deformed fields
$\phi_\iota^\theta$ do not satisfy even the relaxed local
commutativity condition~\cite{ChNST,LS,AG-VM} adapted to this case
and obtained by replacing the light cone with the light wedge.
Nevertheless the fields $\phi_\iota^\theta$ are not completely
delocalized, and because the issues of locality and causality are
crucial for the physical interpretation, the remainder of the
paper is devoted to a precise description of the extent to which
the noncommutative deformation violates locality and local
commutativity.

\section{Wedge locality}\label{S3}

From the definition~\eqref{2.3} of the deformed tensor product, it
directly follows that, for any $f_1, f_2\in S(\oR^4)$ and $g\in
S(\oR^{4n})$,  the following identity holds:
\begin{equation}
(f_1\otimes_\theta(f_2\otimes_{-\theta}g))(x_1,x_2,y)=
(f_2\otimes_{-\theta}(f_1\otimes_\theta g))(x_2,x_1,y),
\label{3.1}
\end{equation}
where $y=(y_1,\dots y_n)$. Indeed, let  $p_1$, $p_2$, and  $q_j$
be the variables conjugate respectively to  $x_1$, $x_2$, and
$y_j$, and let $Q=\sum^n_{j=1}q_j$. The Fourier transform of the
left-hand side of~\eqref{3.1} is $(\hat f_1\otimes\hat
f_2\otimes\hat g)(p_1,p_2,q)$ multiplied by
$\exp\{-\frac{i}{2}(p_1\cdot \theta p_2+p_1\cdot \theta Q-p_2\cdot
\theta Q)\}$, and  that of the right-hand side is obtained by
multiplication with $\exp\{-\frac{i}{2}(-p_2\cdot \theta
p_1-p_2\cdot \theta Q+p_1\cdot \theta Q)\}$. Clearly, these two
multipliers coincide because the matrix  $\theta^{\mu\nu}$ is
antisymmetric.

We will also use the following fact. If  $f\in S(\oR^4)$,  $g\in
S(\oR^{4n})$, and  $f$ has compact support, then we have the
inclusions\footnote{The definition of the Fourier transform used
here is the same as in~\cite{BLOT} and differs from that
in~\cite{GL2} by a sign in exponent, and in consequence the signs
in~\eqref{3.2} differ from those in Eq.~(3.12) of~\cite{GL2}.}
\begin{equation}
\supp(f\otimes_\theta g)\subset \left(\supp f-\tfrac12\theta
U_{\hat g}\right)\times \oR^{4n},\quad \supp(g\otimes_\theta
f)\subset \oR^{4n}\times \left(\supp f+\tfrac12\theta U_{\hat
g}\right), \label{3.2}
\end{equation}
where  $U_{\hat g}$ is the closure of the set $\{Q\in\oR^4\colon
Q=\sum_{j=1}^n q_j,\,\, (q_1,\dots,q_n)\in \supp \hat g\}$.
Indeed,~\eqref{2.3} implies that
\begin{multline}
(f\otimes_\theta g)(x,y)=(2\pi)^{-4(n+1)}\int \hat f(p)\hat g(q)
e^{-ip\cdot  x-i\sum_{j=1}^nq_j\cdot y_j-\frac{i}{2}p\cdot \theta
Q}
dpdq_1\dots dq_n\\
=(2\pi)^{-4n}\int f\left(x+\tfrac12 \theta Q\right)\hat
g(q)e^{-i\sum_{j=1}^nq_j\cdot y_j}dq_1\dots dq_n.
 \label{3*}
\end{multline}
This integral is nonzero only if  $x+\tfrac12
\theta\sum_{j=1}^nq_j$ belongs to $\supp f$ for some
$(q_1,\dots,q_n)\in\supp \hat g$; hence $x\in \supp f-
\tfrac12\theta U_{\hat g}$. The latter set is closed because the
support of $f$ is assumed to be compact. The second inclusion
in~\eqref{3.2} is proved analogously. Now let $f_1,f_2\in
S(\oR^4)$, $g\in S(\oR^{4n})$, $h\in S(\oR^{4m})$, and let $f_1$
and $f_2$ be of compact support. Then it follows from~\eqref{3.2}
that
\begin{equation}
\supp(h\otimes_\theta f_1)\otimes (f_2\otimes_{-\theta}g)\subset
\oR^{4m}\times \left(\supp f_1+\tfrac12\theta U_{\hat
h}\right)\times \left(\supp f_2+\tfrac12\theta U_{\hat
g}\right)\times\oR^{4n}. \label{3.3}
\end{equation}

Following~\cite{GL1,GL2}, we introduce the reference   matrix
\begin{equation}
\theta_1=\begin{pmatrix}
   0&\vartheta_e & 0& 0\\
-\vartheta_e & 0& 0& 0\\
0&0&0&\vartheta_m\\
0&0&-\vartheta_m &0
\end{pmatrix},
\label{3.4}
\end{equation}
where $\vartheta_e\ge0$ and $\vartheta_m\ne 0$, and we let
$\mathrm W_1$ denote the right-hand wedge in Minkowski space,
defined by
\begin{equation}
\mathrm W_1=\{x\in\oR^4\colon x^1>|x^0|\}.
 \label{3.5}
\end{equation}
As shown in~\cite{GL1}, the stabilizer subgroup of the matrix
$\theta_1$ with respect to the action
$\theta\to\Lambda\theta\Lambda^T$ of the proper orthochronous
Lorentz group $L_+^\uparrow$ coincides with that of the wedge
$\mathrm W_1$ with respect to the action $\mathrm
W\to\Lambda\mathrm W$, and there is therefore a one-to-one
correspondence between the orbits of  $\theta_1$ and $\mathrm
W_1$. It is easy to see that if a matrix $\theta$ belongs to the
orbit of $\theta_1$ and $\mathrm W_\theta$ is its corresponding
wedge, then  $-\theta$ also belongs to this orbit and $\mathrm
W_{-\theta}=-\mathrm W_\theta$.

\begin{theorem}\labelT{1}. {\rm (cf. Theorem~4.5 in~\cite{GL2})}  Suppose that $\phi_\iota$ and
$\phi_{\iota'}$ belong to a set of Wightman fields  with the
common domain of definition $D$ spanned by vectors of the
form~\eqref{2.8}. Let $\theta=\Lambda\theta_1\Lambda^T$ and
$\mathrm W_\theta=\Lambda \mathrm W_1$, where $\Lambda\in
L^\uparrow_+$ and where $\theta_1$ and $\mathrm W_1$ are defined,
respectively, by ~\eqref{3.4} and~\eqref{3.5}. If $\phi_\iota$ and
$\phi_{\iota'}$ (anti)commute at spacelike separation,\footnote{As
usual, we assume that the type of commutation relation is the same
for all components of a field.} then the deformed fields
$\phi^\theta_\iota$ and $\phi^\theta_{\iota'}$ satisfy the
(anti)commutation relation
\begin{equation}
[\phi^\theta_\bi(f_1),\phi^{-\theta}_{\bi'}(f_2)]_{\stackrel{-}{(+)}}\,\Phi=0
 \label{3.6}
\end{equation}
for all $\Phi\in D$ and for any $f_1,f_2\in S(\oR^4)$ such that
$\supp f_1\subset \mathrm W_\theta$ and $\supp f_2\subset -\mathrm
W_\theta$.
\end{theorem}

{\it Proof.}  We consider the case when  $\phi_\iota$ and
$\phi_{\iota'}$ commute at spacelike separation. Let
$\Phi_{\bi_1\dots\bi_n} (g)$ be a vector of the form~\eqref{2.8},
where $g\in S(\oR^{4n})$, and let $\Phi_{(m)} (h)$ be a vector of
an analogous form defined by a system of fields
$\phi_{\bi'_1},\dots,\phi_{\bi'_m}$ and  a function $h\in
S(\oR^{4m})$. By the cyclicity   of the vacuum, it suffices to
show that the  assumptions on the supports of $f_1$ and $f_2$
imply that
\begin{equation}
\langle\Phi_{(m)}(h),\,[\phi^\theta_\bi(f_1),\phi^{-\theta}_{\bi'}(f_2)]
\,\Phi_{\bi_1\dots\bi_n} (g)\rangle=0.
 \label{3.7}
\end{equation}
Furthermore, because this matrix element is continuous in  $f_1$
and $f_2$  and smooth functions of compact support are dense in
$S$, we can assume without loss of generality that  $\supp f_1$
and $\supp f_2$ are compact. Using the identity~\eqref{3.1}, this
matrix element can be written as
\begin{equation}
\langle\Phi_{(m)}(h),\,[\phi^\theta_\bi(f_1),\phi^{-\theta}_{\bi'}(f_2)]
\,\Phi_{\bi_1\dots\bi_n} (g)\rangle=(w-w_\pi,\,
h^*\otimes(f_1\otimes_\theta(f_2\otimes_{-\theta}g))),
 \label{3.8}
\end{equation}
where $w_\pi$ is obtained from $w$ by the transposition of the
operators $\phi_\bi(x_1)$ and $\phi_{\bi'}(x_2)$, and where
$h^*(z_1,\dots,z_m)=\overline{h(z_m,\dots,z_1})$.
Using~\eqref{2.10} and the associativity of $\otimes_\theta$, we
obtain
\begin{multline}
(w-w_\pi,\, h^*\otimes(f_1\otimes_\theta(f_2\otimes_{-\theta}g)))=
 (w-w_\pi,\,
h^*\otimes_\theta(f_1\otimes_\theta(f_2\otimes_{-\theta}g)))=\notag\\(w-w_\pi,\,
(h^*\otimes_\theta f_1)\otimes(f_2\otimes_{-\theta}g)).
 \label{3.8}
\end{multline}
Let $k=(k_1,\dots, k_m)$, $p=(p_1, p_2)$, and $q=(q_1,\dots,q_n)$
be the momentum variables conjugate to the coordinates on
$\oR^{4m}\times\oR^{4\cdot2}\times\oR^{4n}$. It follows from the
spectrum condition that
\begin{equation}
\supp(\hat w-\hat
w_\pi)\subset\left\{(k,p,q)\in\oR^{4(m+2+n)}\colon
\sum_{j=1}^mk_j\in \bar V^+,\quad \sum_{j=1}^nq_j\in\bar
V^-\right\}.
 \label{3.9}
\end{equation}
We let  $\bar V^+_\varepsilon$ denote the
$\varepsilon$-neighborhood of the closed forward light cone $\bar
V^+$ and $\chi_\varepsilon$ denote a smoothed characteristic
function  of $\bar V^+$ with the properties: $\chi_\varepsilon$ is
identically 1 on $\bar V^+_{\varepsilon/2}$, vanishes outside
$\bar V^+_\varepsilon$ and is a multiplier of $S(\oR^{4})$.
Because $(u,f)=(2\pi)^{-d}(\hat u,\hat f(-\cdot))$ for any $u\in
S'(\oR^d)$ and $f\in S(\oR^d)$,  it follows from~\eqref{3.9} that
the matrix element~\eqref{3.8} is unchanged on replacing  $h$ and
$g$ with  functions $h_\varepsilon$ and $g_\varepsilon$ such that
$\widehat{h_\varepsilon^*}(k_1,\dots,k_m)=\chi_\varepsilon(-\sum_j
k_j)\widehat{h^*}(k_1,\dots,k_m)$ and $\hat
g_\varepsilon(q_1,\dots,q_n)=\chi_\varepsilon(\sum_j q_j)\hat
g(q_1,\dots,q_n)$. From~\eqref{3.3} we have
\begin{equation}
\supp(h^*_\varepsilon\otimes_\theta f_1)\otimes
(f_2\otimes_{-\theta}g_\varepsilon)\subset \oR^{4m}\times
\left(\supp f_1+\tfrac12\theta \bar V^-_\varepsilon\right)\times
\left(\supp f_2+\tfrac12\theta \bar
V^+_\varepsilon\right)\times\oR^{4n}.
 \notag
\end{equation}
The inclusion $\theta_1V^-\subset \mathrm W_1$ implies that
$\tfrac12\theta V^-\subset\mathrm   W_\theta$ and $\tfrac12\theta
V^+\subset -\mathrm  W_\theta$. Therefore, if the supports of $f_1$
and $f_2$ are compact and contained, respectively,  in $\mathrm
W_\theta$ and  $-\mathrm W_\theta$, and if $\varepsilon$ is
sufficiently small, then  $\supp f_1+\tfrac12\theta \bar
V^-_\varepsilon\subset \mathrm W_\theta$ and $\supp
f_2+\tfrac12\theta \bar V^+_\varepsilon\subset -\mathrm W_\theta$.
Because $(x_1-x_2)^2<0$ for any $x_1\in \mathrm W_\theta$ and
$x_2\in -\mathrm  W_\theta$, we conclude that the
equality~\eqref{3.7} follows from the locality of the undeformed
fields which means, in  terms of the Wightman functions, that
$w-w_\pi$ vanishes for $(x_1-x_2)^2<0$. For the case of
anticommuting fields, the reasoning is the same but with obvious
changes of signs.

\medskip

Because the deformation preserves the translation invariance,
\eqref{3.6} also clearly  holds if there exists a translation $a$
such that
\begin{equation}
\supp f_1+ a\subset\mathrm  W_\theta,\quad \supp
f_2+a\subset-\mathrm W_\theta.
 \notag
\end{equation}

\begin{remark}\label{R1}. {\rm The derivation of Theorem~4.5 in~\cite{GL2}
relies on the assertion that the Fourier transform of the
distribution defined by
$\langle\Psi,\,[\phi^\theta_\bi(f_1),\phi^{-\theta}_{\bi'}(f_2)]
\,\Phi_{\bi_1\dots\bi_n} (g)\rangle$, where $\Psi\in\mathcal H$,
has support in the $(n+2)$-fold product of  the forward light cone
$\bar V^+$. This  contradicts~\eqref{3.6} because then this
distribution would be the boundary value of an analytic function
and hence could not vanish identically on a non-empty open set.
Nevertheless, as shown above, the theorem's conclusion holds.
Another proof of the wedge-local (anti)commutation relations for
the deformed fields is given by Lechner~\cite{L11} in an
operator-algebraic setting.}
\end{remark}

\section{Asymptotic commutativity}\label{S4}

Theorem~\ref{T1} says that the field (anti)commutator
$[\phi^\theta_\bi(x_1),\phi^{-\theta}_{\bi'}(x_2)]_{\stackrel{-}{(+)}}$
vanishes identically on $\mathrm W_\theta\times\mathrm
W_{-\theta}$. In this section, we  show that it also has a rapid
decrease (in the sense of generalized functions) in the whole
spacelike region $(x_1-x_2)^2<0$. As before, we restrict our
consideration to the case of commutator. Let $\Psi$ be an
arbitrary vector in
 $\mathcal H$, let $\Phi$ belong to $D$, and let
\begin{equation}
u_{\Psi,\Phi}(f_1,f_2)=
\langle\Psi,\,[\phi^\theta_\bi(f_1),\phi^{-\theta}_{\bi'}(f_2)]\,\Phi\rangle.
 \label{4.1}
\end{equation}
By the Schwartz kernel theorem, the bilinear
functional~\eqref{4.1} is identified with a distribution in
$S'(\oR^4\times \oR^4)$. A simple way of describing the behavior
of a distribution at infinity is by examining its convolution with
test functions decreasing sufficiently fast. Therefore, we should
consider the asymptotic behavior of the convolution
$u_{\Psi,\Phi}*f$ with adequately chosen functions $f$. This
convolution may be written symbolically in the form
\begin{equation}
(u_{\Psi,\Phi}*f)(x_1,x_2)=\int\langle\Psi,\,
[\phi^\theta_\bi(\xi_1),\phi^{-\theta}_{\bi'}(\xi_2)]\,\Phi\rangle
f(x_1-\xi_1,x_2-\xi_2) \,d\xi_1 d\xi_2.
 \label{4.2}
\end{equation}

We will use the Gelfand-Shilov test function spaces
$S^{\beta}_{\alpha}$ which are contained in $S$. If $\beta<1$, the
definition of these spaces can be formulated in terms of complex
variables, which considerably simplifies the estimates of
Theorem~\ref{T2} below. As shown in~\cite{GS3}, the elements of
$S^{\beta}_{\alpha}(\oR^d)$, where $\beta<1$,  can be continued
analytically into $\oC^d$, and $S^{\beta}_{\alpha}$ is isomorphic
to the space of entire functions  $W^\sigma_\rho$, where
$\rho=1/\alpha$ and $\sigma=1/(1-\beta)>1$. The functions
belonging to $W^\sigma_\rho$ satisfy the inequality
\begin{equation}
|f(x+iy)|\le C\prod_{j=1}^de^{-a|x_j|^\rho+b|y_j|^\sigma}
 \label{4.3}
\end{equation}
with some positive constants $a$, $b$, and $C$ depending on  $f$.
The norm corresponding to~\eqref{4.3} is given by
\begin{equation}
\|f\|_{a, b}=\sup_{z=x+iy}|f(z)|\prod_{j=1}^d e^{a|x_j|^\rho-
b|y_j|^\sigma} \label{4.4}
\end{equation}
and we let  $W^{\sigma, b}_{\rho,a}$ denote the space of entire
functions such that $\|f\|_{\bar a,\bar b}<\infty$ for all
positive $\bar a<a$ and $\bar b>b$. Clearly,
\begin{equation}
W^\rho_\gamma=\bigcup_{a\to0,\, b\to\infty}W^{\rho, b}_{\gamma,a}.
\notag
\end{equation}
 If $\sigma>\rho$, then the space $W_{\rho, a}^{\sigma,b}$
is nontrivial for any  $a>0$ and $b>0$, but if $\sigma=\rho$,
$W_{\rho, a}^{\sigma,b}$ is nontrivial only  under the condition
$a\ge b$. Indeed, if $a<b$, then~\eqref{4.3} implies that
$f(z)\cdot f(iz)$ tends to zero as  $|z|\to\infty$ and is hence
identically zero by the Liouville theorem. The same argument shows
that $W_\rho^\sigma$ is trivial for $\sigma<\rho$. Under the
condition $\rho>1$, the Fourier transformation is an isomorphism
of $W^{\sigma, b}_{\rho,a}$ onto $W_{\sigma', b'}^{\rho',a'}$,
where  the primed indices are defined by the duality relations
\begin{equation}
\frac{1}{\rho'}+\frac{1}{\rho}=1,\qquad (\rho'a')^\rho(\rho
a)^{\rho'}=1,
 \label{4.5}
\end{equation}
and by analogous relations for $\sigma'$, $b'$. We will also use
the spaces $W^{\sigma, b}$ defined by
\begin{equation}
|f(z)|\le C_{N,\bar b}(1+|x|)^{-N}\prod_{j=1}^de^{\bar
b|y_j|^\sigma},\quad \bar b>b,\,\, N=0,1,2\dots,
 \label{**}
\end{equation}
and the spaces  $W^\sigma=\bigcup_{b\to\infty}W^{\sigma, b}$,
which are isomorphic to the Gelfand-Shilov spaces $S^\beta$ with
$\beta=1-1/\sigma$. The Fourier  transformation  maps  $W^\sigma$
onto the space $W_{\sigma'}=S_{1/\sigma'}$ and  $W^{\sigma, b}$
onto the space $W_{\sigma',b'}$ of smooth functions on  $\oR^d$
with the  norms
\begin{equation}
\|g\|_{N,\bar b'}=\max_{|\kappa|<N}\sup_p\prod_{j=1}^de^{\bar
b'|p_j|^{\sigma'}}|\partial^\kappa g(p)|,\qquad \bar b'<b', \quad
N=0,1,2,\dots.
 \notag
\end{equation}
The choice of norm $|\cdot|$ on $\oR^d$ is inessential  to the
definition~\eqref{**}, but  when working with functions in
$W^\sigma_\rho$,  it is convenient to use the norm
\begin{equation}
|x|=\left(\sum_j|x_j|^\rho\right)^{1/\rho}.
 \label{4.6}
\end{equation}
We need two auxiliary lemmas.

\begin{lemma}\label{L1}. Let $u$ be a distribution on $\oR^d$ with support in a
closed cone $V\ne\oR^d$, and let $G$ be a closed cone such that
$G\cap V=\{0\}$. If $f\in W_{\rho,a}$, then for some $N$ and for
any $\Bar{\Bar a}<\Bar a<a$, the convolution $(u*f)(x)$ satisfies
the estimate
\begin{equation}
|(u*f)(x)|\le C_{G,\bar a,\Bar{\Bar a}}\|f\|_{N,\bar a}
e^{-\Bar{\Bar a}|d_{{}_{G,V}}\, x|^\rho} ,\qquad  x\in G,
 \label{4.7}
\end{equation}
where the norm $|\cdot|$ on $\oR^{d}$ is given by~\eqref{4.6} and
$d_{G,V}=\inf_{x\in G, |x|=1}\inf_{\xi\in V}|x-\xi|$.
\end{lemma}

{\it Proof.}  For simplicity we assume that the set $V$ is
regular.\footnote{See definition in~\cite{BLOT}, Supplement A.2.
The closed light cone is a regular set.} Then there exist a
constant    $C>0$ and a integer  $N$ (both depending on $u$) such
that
\begin{equation}
|(u,f)|\le C\max_{|\kappa|\le N}\,\sup_{\xi\in V}
(1+|\xi|)^N|\partial^\kappa f(\xi)|\quad \text{for all $f\in
S(\oR^d)$}. \label{4.8}
\end{equation}
If the regularity condition is not satisfied, then  $V$
in~\eqref{4.8} should be replaced by its
$\varepsilon$-neighborhood. This slightly complicates the
analysis, but does not change the result. By the definition of
norms in $W_{\rho,a}$, we have $|\partial^\kappa
f(\xi)|\le\|f\|_{|\kappa|,\bar a}e^{-\bar a|\xi|^\rho}$ for any
$\kappa$ and $\bar a<a$. Replacing the function $f(\xi)$  by
$f(x-\xi)$ and using~\eqref{4.8},  we obtain
\begin{multline}
|(u*f)(x)|\le C\|f\|_{N,\bar a}
\sup_{\xi\in V}(1+|\xi|)^N e^{-\bar a|x-\xi|^\rho}\le\\
\le C\|f\|_{N,\bar a}(1+|x|)^N \sup_{\xi\in V}(1+|x-\xi|)^N
e^{-\bar a|x-\xi|^\rho}\le\\
\le  C_{\bar a_1} \|f\|_{N,\bar a} (1+|x|)^N\,e^{-\bar
a_1|d_V(x)|^\rho}
  \label{4.10}
\end{multline}
where $\bar a_1<\bar a$ and can be chosen arbitrarily close to
$\bar a$, and $d_V(x)=\inf_{\xi\in V}|x-\xi|$. Because the cone
$V$ is invariant under dilations, we have
\begin{equation}
d_V(x)=|x|\inf_{\xi\in V}|x/|x|-\xi|=|x|d_V(x/|x|). \notag
\end{equation}
It follows from the condition $G\cap V=\{0\}$ that $d_{G,V}=\inf_{x\in G,
|x|=1}d_V(x)>0$. Therefore the factor  $(1+|x|)^N$ can be omitted
from the last row in~\eqref{4.10}, slightly decreasing  $\bar
a_1$, and we arrive at~\eqref{4.7}. Lemma~\ref{L1} is proved.

\begin{remark}\label{R2}. {\rm We consider below a special case, where $d=d_1+d_2$ and
$\supp u\subset \oV\times \oR^{d_2}$, with $\oV$   a cone in
$\oR^{d_1}$. Then an estimate analogous to~\eqref{4.7} holds in
any closed cone $\oG\subset \oR^{d_1}$ such that $\oG\cap
\oV=\{0\}$ and even under a weaker assumption on the behavior of
$f$ with respect to the second group of variables. In particular,
those functions are admissible that satisfy the conditions
\begin{equation}
\max_{|\kappa|\le N}|\partial^\kappa f(x, x')|\le C_{f,N,\bar
a}e^{-\bar a|x|^\rho}(1+|x'|)^{-N},\quad
 N=0,1,2,\dots.
 \label{4.11}
\end{equation}
From~\eqref{4.11}, we obtain an estimate of type~\eqref{4.7} for
$(u*f)(x,0)$, but with $C_{f,N,\bar
a}$ in place of $\|f\|_{N,\bar a}$. The function space defined by~\eqref{4.11} is the completed
tensor product $W_{\rho,a}(\oR^{d_1}) \hat{\otimes}S(\oR^{d_2})$.}
\end{remark}

\begin{lemma}\label{L2}.  If  $\sigma'>\rho=\rho'/(\rho'-1)$, then for
every quadratic form  $\mathcal Q(p)$ with real coefficients,
 the function  $e^{i\mathcal Q(p)}$ is a multiplier of
$W^{\rho',a}_{\sigma',b}$  for any  $a>0$, $b>0$.
\end{lemma}

{\it Proof.} We need to estimate the function  $|e^{i\mathcal
Q(p+is)}|=e^{-\Im \mathcal Q(p+is)}$. Let $\mathcal Q_{jk}$ be the
matrix of the quadratic form $\mathcal Q$ and let $|\mathcal
Q|=\max_{j,k}|Q_{jk}|$.   Young's inequality for products states
that if $r$ and $t$ are nonnegative real numbers and $\rho$ and
$\rho'$ are positive numbers satisfying the first of duality
relations~\eqref{4.5}, then $rt\le r^\rho/\rho+t^{\rho'}/\rho'$.
Using this inequality with $r=|p_j|/\varepsilon$ and
$t=\varepsilon|s_k|$, where $\varepsilon>0$, we obtain
\begin{multline}
\left|\,\Im\sum_{j,k}(p_j+is_j)\mathcal Q_{jk}(p_k+is_k)\right|\le
2|\mathcal Q|\sum_{j.k}|p_js_k|\le\\
\le 2d|\mathcal
Q|\sum_j\left(\frac{1}{\rho}\left|\frac{p_j}{\varepsilon}
\right|^\rho+\frac{1}{\rho'}|\varepsilon s_j|^{\rho'}\right).
 \label{4.12}
\end{multline}
The condition $\rho<\sigma'$ implies that for  arbitrarily small
$\varepsilon$, there is a constant $C_\varepsilon>0$ such that
\begin{equation}
|p_j/\varepsilon|^\rho\le
C_\varepsilon+\varepsilon|p_j|^{\sigma'}.
 \label{4.13}
\end{equation}
In line with~\eqref{4.4}, the norms in $W^{\rho'}_{\sigma'}$ are
defined by $\|g\|_{a, b}=\sup_{p,s}|g(p+is)|\prod_{j=1}^d
e^{a|p_j|^{\sigma'}- b|s_j|^{\rho'}}$. Substituting~\eqref{4.13}
into~\eqref{4.12}, we conclude that for any $\bar a_1<\bar a<a$
and $\bar b_1>\bar b>b$, there exists a constant $C_{\bar a_1,\bar
b_1}$ such that
\begin{equation}
\|ge^{i\mathcal Q}\|_{\bar a_1,\bar b_1}\le C_{\bar a_1,\bar b_1}
\|g\|_{\bar a,\bar b}\qquad\text{for all $g\in
W^{\rho',a}_{\sigma',b}$},
 \notag
\end{equation}
which completes the proof.

\begin{remark}\label{R3}. {\rm If $\sigma'=\rho$, then $e^{iQ(p)}$ is a
multiplier of $W^{\rho'}_{\sigma'}$, but not of
$W^{\rho',a}_{\sigma',b}$. The condition $\sigma'\ge\rho$ for the
spaces of type $W$ is equivalent  to the condition
$\alpha\ge\beta$ for $S^\beta_\alpha$. As shown in~\cite{S07TMP},
only under this condition  $S^\beta_\alpha$ is an algebra with
respect to the Weyl-Moyal product~\eqref{1.4}.}
\end{remark}

We now turn to describing the asymptotic behavior of
distribution~\eqref{4.1} at large spacelike separations. The
corresponding theorem is accompanied below by a simple but
explanatory example. We let $\oV$ denote the cone in
$\oR^{4\cdot2}=\oR^4\times\oR^4$ consisting of the pairs
$(x_1,x_2)$ such that $x_1-x_2$ belongs to the closed light cone,
\begin{equation}
\oV=\{(x_1,x_2)\in\oR^{4\cdot 2}\colon (x_1-x_2)^2\ge0\}.
 \notag
\end{equation}

\begin{theorem}\label{T2}.  Let $\Phi(g)$ be a vector of the
form~\eqref{2.8} with $g\in W^\sigma(\oR^{4n})$, let $\Psi$ be an
arbitrary vector in $\mathcal H$, and let $\oG$  be a closed cone
in $\oR^{4\cdot2}$ such that $\oG\cap \oV=\{0\}$. If $f\in
W^{\sigma}_{\rho}(\oR^{4\cdot 2})$, where
$\rho<\sigma'=\sigma/(\sigma-1)$ and $\|f\|_{a,b}<\infty$, then
for any $\bar a<a$, the function~\eqref{4.2} satisfies the
estimate
\begin{equation}
|(u_{\Psi,\Phi}*f)(x_1,x_2)|\le C_{G,\bar a\Psi,\Phi}\|f\|_{a,b}
\exp\left\{-\Bar{a}\,
d^\rho_{\oG,\oV}\left(|x_1|^\rho+|x_2|^\rho\right)\right\},\quad
(x_1,x_2)\in \oG,
 \label{4.14}
\end{equation}
where the angular distance $d_{\oG,\oV}$ is defined in
Lemma~\ref{L1}.
\end{theorem}

{\it Proof.} Let $u_\Psi$ be the distribution defined on
$\oR^{4(2+n)}$ by the three-linear functional
$\langle\Psi,\,[\phi_\bi(f_1),\phi_{\bi'}(f_2)]\,\Phi (g)\rangle$.
By locality of the undeformed fields, its support lies in the cone
$\oV\times\oR^{4n}$, and~\eqref{3.1} implies that
\begin{equation} (u_{\Psi,\Phi},f_1\otimes f_2)=
(u_\Psi,f_1\otimes_\theta(f_2\otimes_{-\theta}g))\quad\text{for
all $f_1, f_2\in S(\oR^4)$ and $g\in S(\oR^{4n})$}.
 \label{4.15}
\end{equation}
By the definition of the product $\otimes_\theta$, the Fourier
transform of $f_1\otimes_\theta(f_2\otimes_{-\theta}g)$ has the
form $\eta\cdot(\hat f_1\otimes\hat f_2\otimes\hat g)$, where
\begin{equation}
\eta(p_1,p_2,q)=e^{-\tfrac{i}{2}\left(p_1\theta
p_2+(p_1-p_2)\theta\sum^n_{j=1}q_j\right)}. \label{4.16}
\end{equation}
Since $S(\oR^4)\otimes S(\oR^4)$ is dense in $S(\oR^{4\cdot2})$,
it follows from~\eqref{4.15} that for any $f\in S(\oR^{4\cdot2})$,
\begin{equation}
(u_{\Psi,\Phi},f)=(u_\Psi,h_f), \quad \text{where}\,\,\hat
h_f\eqdef \eta\cdot(\hat f\otimes \hat g)\in S(\oR^{4(2+n)}).
 \notag
\end{equation}
Therefore,
\begin{equation}
(u_{\Psi,\Phi}*f)(x_1, x_2)=(u_\Psi*h_f)(x_1,x_2,0).
 \label{4.17}
\end{equation}
If $f\in W^\sigma_\rho(\oR^{4\cdot 2})$ and $g\in
W^\sigma_\rho(\oR^{4n})$, then  $\|f\otimes g\|_{a,b}=
\|f\|_{a,b}\|g\|_{a,b}$ by the definition~\eqref{4.4}.  The
Fourier transformation, as already said, is an isomorphism  of
$W^{\sigma, b}_{\rho,a}$ onto $W_{\sigma', b'}^{\rho',a'}$, and by
Lemma~\ref{L2} the function $\eta$ is a multiplier of $W_{\sigma',
b'}^{\rho',a'}(\oR^{4(n+2)})$ under the condition $\rho<\sigma'$.
Hence the correspondence $f\to h_f$ is continuous from
$W^{\sigma,b}_{\rho,a}(\oR^{4\cdot2})$ to
$W^{\sigma,b}_{\rho,a}(\oR^{4(n+2)})$, and for any $\bar a_1<\bar
a<a$  and $\bar b_1>\bar b>b$, we have
\begin{equation}
\|h_f\|_{\bar a_1,\bar b_1}\le C_{\bar a_1,\bar b_1} \|f\|_{\bar
a,\bar b}\le  C_{\bar a_1,\bar b_1} \|f\|_{a, b}.
  \label{4.18}
\end{equation}
The operation of differentiation  is continuous in
$W^{\sigma,b}_{\rho,a}$,   as can easily  be seen by using
Cauchy's formula. Therefore,  $W_{\rho, a}$ is continuously
embedded in $W^{\sigma,b}_{\rho,a}$, and
\begin{equation}
\|h_f\|_{N,\bar a_2}\le C_{\kappa,\bar a_2} \|h_f\|_{\bar a_1,\bar
b_1}\quad\text{for any $N$ and $\bar a_2< \bar a_1$}.
  \label{4.19}
\end{equation}
Applying Lemma~\ref{L1} to the right-hand side of~\eqref{4.17} and
using~\eqref{4.18} and \eqref{4.19}, we arrive at~\eqref{4.14}.
Now let $g\in W^{\sigma,B}$. Performing the Fourier
transformation, using the condition $\rho<\sigma'$, and making the
inverse transformation, we obtain
\begin{equation}
|h_f(x+iy,x'+iy')|\le C_{g,\bar a,\bar b,\bar
B}\|f\|_{a,b}e^{-\bar a|x|^\rho}(1+|x'|)^{-N} \prod_{j,\mu}e^{\bar
b|y_j^\mu|^\sigma+ \bar B|y'^\mu_j|^\sigma}.
  \notag
\end{equation}
where $x$    denotes the pair $(x_1,x_2)$ and $x'$ denotes the
variables of $g$. Hence $h_f(x,x')$ satisfies inequalities of
type~\eqref{4.11} with a constant $C_{f,N,\bar a}$ proportional to
$\|f\|_{a,b}$. Invoking Remark~\ref{R2}, we arrive again
at~\eqref{4.14}, which completes the proof.

\begin{remark}\label{R4}. {\rm The condition $\oG\cap \oV=\{0\}$ implies that
$\oG$ is contained in a wedge of the form $\{(x_1,x_2)\colon
x_1-x_2\in G\}$, where $G$ is a closed cone in $\oR^4$ having only
the origin in common with the closed light cone $\bar V$. The
norm~\eqref{4.6} dominates the Euclidean norm $\|\cdot\|$ and
using the parallelogram identity, we see that for all $x_1-x_2\in
G$, the following inequality holds:
\begin{equation}
|(u_{\Psi,\Phi}*f)(x_1,x_2)|\le C_{G,\Psi,\Phi}\|f\|_{a,b}
\exp\left\{-\frac{a}{2}\delta_{G,\bar V}\|x_1-x_2\|^\rho\right\},
 \notag
\end{equation}
where the distance $\delta_{G,\bar V}$ is defined by the Euclidean
norm.}
\end{remark}

As an explanatory example,  consider the matrix element
\begin{equation}
u_{g_1,g_2}(f_1,f_2)=\langle\varphi(g_1)\,\Omega,\,
[\varphi^\theta(f_1),\varphi^{-\theta}(f_2)]
\,\varphi(g_2)\,\Omega\rangle,
 \label{4*}
\end{equation}
where $\varphi$ is a  free massive neutral scalar field.
Expressing the four-point vacuum expectation value in terms of
two-point ones and passing to the momentum representation,
\eqref{4*} can be written as
\begin{multline}
u_{g_1,g_2}(f_1,f_2)=\frac{i}{(2\pi)^8}\int dk dp_1 dp_2
dq\,\delta(k+q)\delta(p_1+p_2)\hat \varDelta_+(k)\hat
\varDelta(p_1)
\\
\times e^{-\tfrac{i}{2}(p_1\theta p_2+p_1\theta q-p_2\theta q)}\,
\overline{\hat g_1(k)} \hat f_1(-p_1)\hat f_2(-p_2) \hat g_2(-q)
\\
= \frac{i}{(2\pi)^8}\int dk dp \, \hat \varDelta_+(k)\hat
\varDelta(p)\, e^{ip\theta k}\,\overline{\hat g_1(k)} \hat
f_1(-p)\hat f_2(p) \hat g_2(k),
 \notag
\end{multline}
where $\hat\varDelta(p)=-2\pi i\epsilon(p^0)\delta(p^2-m^2)$ is
the Fourier transform of the Pauli-Jordan function, and
$\hat\varDelta_+$ is its positive-frequency part. Letting $\hat
f(p)=\hat f_1(p)\hat f_2(-p)$ and $\hat g(k)=\overline{\hat
g_1(-k)}\hat g_2(-k)$ and turning back to the coordinate
representation, we obtain
\begin{equation}
u_{g_1,g_2}(f_1,f_2)=i\left(\varDelta_+\otimes_{2\theta}\varDelta,
g\otimes f\right) =i\left(\varDelta_+\otimes\varDelta,
g\otimes_{2\theta} f\right).
 \label{4**}
\end{equation}
If $\supp f_1\subset \mathrm W_\theta$ and $\supp f_2\subset
\mathrm W_{-\theta}$, then $f$, being the convolution product of
$f_1(\xi)$ and $f_2(-\xi)$, is supported in $\mathrm W_\theta$.
Taking into consideration the  support properties of $\hat
\varDelta_+$, we conclude, as in Sec.~\ref{S3},  that
$\supp(g\otimes_{2\theta} f)\subset \mathrm W_\theta+\theta
V^-\subset \mathrm W_\theta$, and hence $u_{g_1,g_2}(f_1,f_2)$
vanishes for such test functions. To test the behavior
of~\eqref{4*} for arbitrary  spacelike separations, we use the
shifted test functions  $f_1(x_1-\xi)$ and $f_2(x_2-\xi)$, where
$(x_1-x_2)^2<0$. Then $f(\xi)$ is replaced by $f(x_1-x_2-\xi)$ and
$g\otimes_{2\theta} f$ is shifted away from the support of
$\varDelta_+\otimes\varDelta$. Therefore,
$u_{g_1,g_2}(f_1(x_1-\cdot),f_2(x_2-\cdot))$ inherits the fall-off
properties of $g\otimes_{2\theta} f$. It is clear from~\eqref{3*}
that the rate of decrease of $g\otimes_{2\theta} f$ in the
$x$-variable is the same as that of $f$,  if the latter decreases
slower than $\hat g$ does. In technical terms, if $\hat g_2(k)$,
and hence $\hat g(k)$, belongs to $W_{\sigma'}=\widehat{W}^\sigma$
and falls off as $e^{-|k|^{\sigma'}}$,  we take $f_{1,2}$ in
$W_{\rho}$ with $\rho<\sigma'$ and conclude that
$u_{g_1,g_2}(f_1(x_1-\cdot),f_2(x_2-\cdot))$ decreases  with a
rate characterized  by Lemma~1.  In the  case of interacting
fields, the occurrence of the term $-\tfrac{i}{2}p_1\theta p_2$
in~\eqref{4.16} forces us to take $f_1$ and $f_2$  in $W_\rho\cap
W^\sigma=W_{\rho}^\sigma$. With such a choice, the
function~\eqref{4.2} falls off in the same manner in all spacelike
directions.

It should be pointed out  that the linear subspace spanned by the
vacuum $\Omega$ and all vectors of the form~\eqref{2.8} with $g\in
W^\sigma$ is dense in $\mathcal H$, because $W^\sigma$ is dense in
$S$. Any   $W^\sigma$ contains $W^1$  whose Fourier transform is
nothing but the space $C^\infty_0$ of all infinitely
differentiable functions of compact support, and $C^\infty_0$ is
just the function space that is employed in the Haag-Ruelle
scattering theory. The condition $\rho<\sigma'$, together with the
condition $\rho\le\sigma$ of non-triviality of $W^\sigma_\rho$,
implies that $\rho<2$, because $\min(\sigma,\sigma')\le2$.  Since
$\rho$ can be chosen arbitrarily close to~2,  Theorem~\ref{T2}
merely says  that the commutator
$[\phi^\theta_\bi(x_1),\phi^{-\theta}_{\bi'}(x_2)]$ decreases
approximately as a Gaussian at large spacelike separation of $x_1$
and $x_2$. The borderline case $\rho=\sigma$, i.e., the case of
test functions in $W^\sigma_\sigma=S^\beta_{1-\beta}$, where
$\beta=1-1/\sigma=1/\sigma'$, is of particular
interest.\footnote{We note that $\rho<\sigma'$ implies
$\beta<1/2$.} Theorem~3 of~\cite{FS} shows, that in this case,
\eqref{4.14} amounts to  the condition that the distribution
$u_{\Psi,\Phi}$ has a continuous extension to the space
$S^\beta(\oV)=W^\sigma(\oV)$ of entire functions satisfying the
inequalities
\begin{equation}
|f(x+iy)|\le C_N (1+ |x|)^{-N} \exp\left\{b
d^\sigma_U(x)+b|y|^\sigma\right\},\qquad N=0,1,2,\dots,
 \label{4.20}
\end{equation}
where $U$ is an open cone, depending on $f$, such that
$\oV\setminus \{0\}\subset U$, and where $C_N$ and $b$ are
positive constants, also depending on $f$. Conditions of this kind
were earlier used in nonlocal quantum field theory, where the
framework of tempered distributions appears to be too restrictive
and the adequate choice of test function space takes on great
significance. In particular, the spaces $S^\beta(\oV)$ with
$\beta<1$, consisting of analytic functions,  were used in
formulating  an asymptotic commutativity principle replacing local
commutativity  for nonlocal fields. In~\cite{FS}, nonlocal fields
$\phi_\iota$ and $\phi_{\iota'}$ defined as operator-valued
generalized functions on $S^\beta(\oR^4)$ are referred to as
asymptotically (anti)commuting, if the matrix element
$\langle\Psi,[\phi_\iota(x_1),\phi_{\iota'}(x_2)]_{\stackrel{-}{(+)}}
\Phi\rangle$ has a continuous extension to  $S^\beta(\oV)$ for any
vectors $\Phi$ and $\Psi$ in their common dense domain in the
Hilbert space. The principle of asymptotic commutativity implies
that any two field components either commute or anticommute
asymptotically at large spacelike separation of the arguments.
This condition provides a way of extending the CPT and
spin-statistics theorems to nonlocal QFT~\cite{S99}.  The
condition of mutual asymptotic commutativity was also used
in~\cite{S01} to extend the Borchers equivalence classes  to
nonlocal fields. Theorem~\ref{T2} shows that the deformed fields
$\phi^\theta_\iota$ and $\phi^{-\theta}_{\iota'}$ (more precisely,
their restrictions to the test functions in $S^\beta(\oR^4)$,
$\beta<1/2$)  (anti)commute asymptotically if the initial fields
$\phi_\iota$ and $\phi_{\iota'}$ (anti)commute at spacelike
separation. Using the fact that the twist operator~\eqref{1.3} is
an automorphism of $S^\beta(\oV)$ for $\beta<1/2$, an extension of
the distribution
$\langle\Psi,[\phi^\theta_\iota(x_1),\phi^{-\theta}_{\iota'}(x_2)]_{\stackrel{-}{(+)}}
\Phi\rangle$ to this space can explicitly be constructed, but here
we find it preferable to define the asymptotic commutativity as a
fall-off property of the smoothed field commutator, which clearly
shows its meaning.

\section{Cluster properties of the deformed Wightman functions}\label{S5}

In order to prove the uniqueness of the vacuum state in
reconstructing quantum fields from a given set of Wightman
functions, it suffices to use the cluster decomposition property
\begin{equation}
w_{\bi_1\dots\bi_n}(f\otimes g_{(\lambda a)})\longrightarrow
w_{\bi_1\dots\bi_m}(f)w_{\bi_{m+1}\dots\bi_n}(g)  \qquad
(\lambda\to \infty),
 \notag
\end{equation}
where $f\in S(\oR^{4m})$, $g\in S(\oR^{4(n-m)})$, and $a$ is an
arbitrary  spacelike vector. However, Theorem~3-4 of~\cite{SW}
shows that the vacuum expectation values of local field theory
satisfy the slightly stronger condition
\begin{equation}
w_{\bi_1\dots\bi_n}(h_{(m,\lambda a)})\longrightarrow
(w_{\bi_1\dots\bi_m}\otimes w_{\bi_{m+1}\dots\bi_n})(h)  \qquad
(\lambda\to \infty),
 \label{5.2}
\end{equation}
where $h$ is any function in $S(\oR^{4n})$ and
\begin{equation}h_{(m,\lambda
a)}(x_1,\dots,x_n)=h(x_1,\dots,x_m,x_{m+1}-\lambda
a,\dots,x_n-\lambda a).
 \notag
\end{equation}
Setting
\begin{equation}
\hat h=\prod_{1\le j<k\le n}e^{-\frac{i}{2}p_j\theta p_k}(\hat
f\otimes \hat g)
 \notag
 \end{equation}
and using~\eqref{2.3} and~\eqref{2.6}, we see  that the limit
relation~\eqref{5.2} implies~\eqref{2.7} for the deformed Wightman
functions, because the distribution $w_{\bi_1\dots\bi_m}\otimes
w_{\bi_{m+1}\dots\bi_n}$ contains the factor
$\delta\left(\sum^m_{j=1}p_j\right)\delta\left(\sum^n_{j=m+1}p_j\right)$
by the translation invariance.

The Haag-Ruelle scattering theory  uses essentially the
decomposition of vacuum expectation values into truncated ones.
The truncated Wightman functions $w^T$ are obtained by
eliminating  the contribution of the intermediate vacuum state
from the support of $\hat w$, see~\cite{BLOT,J}. If zero is an
isolated point of the spectrum of the energy-momentum operator,
i.e., the spectrum has the form
\begin{equation}
\spec P\subset \{0\}\cup \bar V_\mu^+,\qquad \text{where $\bar
V_\mu^+=\{p\colon p_0\ge\sqrt{\mathbf p^2+\mu^2}\}$ and $\mu>0$},
 \label{5.4}
\end{equation}
then  $\supp \hat w^T(p_1,\dots,p_n)$  is contained in the set
defined by
\begin{equation}
\sum_{j=1}^n p_j=0,\qquad \sum_{j=1}^kp_j\in \bar V_\mu^+,\quad
k=1,\dots,n-1.
 \notag
\end{equation}
The asymptotic behavior  of the truncated vacuum expectation
values at spacelike infinity plays the major role in constructing
the  scattering states. Because of this, it is desirable to
elucidate how  the deformation under consideration affects this
behavior. In accordance with~\eqref{2.6}, we define the deformed
truncated $n$-point  vacuum expectation values by
\begin{equation}
w^{T,\bt}_{\,\bi_1\dots\bi_n}(f_1\otimes \dots\otimes f_n)\eqdef
w^T_{\,\bi_1\dots\bi_n}(f_1\otimes_\theta\dots\otimes_\theta f_n).
 \label{5.5}
\end{equation}
The usual method~\cite{BLOT, J} of  estimating the spacelike
asymptotic behavior of $w^T$, as well as the proof of  cluster
property~\eqref{5.2} in~\cite{SW}, is based on Ruelle's auxiliary
theorem which can be given the following form:

If two tempered distributions  $u_1$ and $u_2$ coincide on an open
cone $\Gamma$ and the supports  of their Fourier transforms are
separated by a finite distance, then both of these distributions
vanishes at infinity  faster than any inverse power of $|x|$ in
any closed cone $G$ such that $G\setminus\{0\}\subset \Gamma$.

Indeed, for each  test function  $f\in S$, the convolution
$(u_1-u_2)*f$ together with all its derivatives decreases rapidly
in any direction within  $\Gamma$, because a shift inside this
cone implies that the test function moves away from support of
$(u_1-u_2)$. The corresponding estimate is similar to that made in
the proof of Lemma~\ref{L1} for the case of test functions in
$W^\sigma_\rho\subset S$.  Let now $\chi(p)$  be a multiplier of
$S$, equal to 1 on a neighborhood of $\supp\hat u_1$ and equal to
zero on a neighborhood of $\supp\hat u_2$. Then we have the
identity
\begin{equation}
u_1*f=(u_1-u_2)*(\chi*f),
 \label{5.6}
\end{equation}
 which shows that $u_1*f$ also rapidly
decreases inside $\Gamma$.

This  theorem is applied to the truncated vacuum expectation
values in the following way.  Let $J$ be a nonempty subset of the
set of indices $(1,2,\dots, n)$ with a nonempty complement $J'$.
We let $\pi$ denote the permutation  $(1,2,\dots, n)\to (J,J')$
and $\pi'$ denote the permutation $(1,2,\dots, n) \to (J',J)$. By
local commutativity,  $w^T$ coincides with  the permuted
distribution $w^T_\pi$ on the cone
\begin{equation}
\Gamma_J=\bigcap_{j\in J,j'\in
J'}\Gamma_{jj'},\qquad\mbox{where}\quad
\Gamma_{jj'}=\{x\in\oR^{4n}\colon (x_j-x_{j'})^2<0\},
 \notag
\end{equation}
and $w^T_\pi$ in turn coincides with  ñ $w^T_{\pi'}$ on this cone.
It follows from the spectrum condition that $\hat w^T_{\pi'}=0$ if
$\sum_{j\in J}p_j=P_J\notin V^+_\mu$, and that $\hat w^T_{\pi}=0$
if $P_J\notin V^-_\mu$, because $P_J+P_{J'}=0$ by the translation
invariance. The cones $\Gamma_J$ with various $J$ cover the plane
$x_1^0=\dots=x_n^0$ in $\oR^{4n}$, and the Ruelle theorem says
that, for any $f\in S(\oR^{4n})$, the function
$w^T_{\,\bi_1\dots\bi_n}*f$ restricted to this plane and
considered as a function of the difference variables $\mathbf
x_j-\mathbf x_{j+1}$ belongs to $S(\oR^{3(n-1)})$. It is precisely
the property of $w^T_{\,\bi_1\dots\bi_n}*f$ that is used
in~\cite{BLOT, J} to prove the existence of  asymptotic scattering
states.

The Ruelle theorem can be considerably strengthened  using the
freedom in choosing the multiplier  $\chi$ in~\eqref{5.6}. By
varying this multiplier,  the distributions $u_1$ and $u_2$ can be
shown to decay exponentially inside $G$ with a rate constant
determined by the distance between  $\supp \hat u_1$ and $\supp
\hat u_2$. In~\cite{S82}, this improvement is reduced to an
extremum problem whose solution is expressed  through Chebyshev
polynomials. To detect this decay,  appropriate test functions are
needed. Clearly, they should decrease sufficiently fast at
infinity, and we use the functions in $S$ that have exponential
decrease of order $\ge 1$ and  type $\ge 1/l$, and   satisfy the
condition
\begin{equation}
\|f\|_{N,\bar l}\eqdef\max_{|\kappa|\le N}\,\sup_x|\partial^\kappa
f(x)|\prod_je^{|x_j|/\bar l}<\infty
 \notag
\end{equation}
for all $\bar l>l$ and $N=0,1,\dots$. Letting $S_{1,l}$ denote
this space, the result~\cite{S82} can be stated as follows. In
local quantum field theory with the spectrum
condition~\eqref{5.4}, the convolution of
$w^T_{\,\bi_1\dots\bi_n}$ with any test function $f\in S_{1,l}$
satisfies the estimate
\begin{equation}
\left.|\partial^\kappa
(w^T_{\,\bi_1\dots\bi_n}*f)(x)|\right|_{x_1^0=\dots =x_n^0}\le
C_{\bar l}\,\|f\|_{|\kappa|+K,\bar l} \exp\left\{-\frac{\mu
R}{2(n-1)(1+3\mu\bar l)}\right\},
 \label{5.7}
\end{equation}
where $\bar l>l$ and can be taken arbitrarily close to $l$, the
constant $K$ is determined by the order of singularity of
$w^T_{\,\bi_1\dots\bi_n}$, and
\begin{equation}
R=\max_{j,k}\|\mathbf x_j-\mathbf x_k\| .
 \notag
\end{equation}
With $l\ll 1/\mu$, \eqref{5.7} shows that
$w^T_{\,\bi_1\dots\bi_n}$ decays  no slower than $\exp\{-\mu
R/2(n-1)\}$ as $R\to\infty$. In order to characterize the behavior
of the deformed functions $w^{T,\theta}_{\,\bi_1\dots\bi_n}$ at
infinity, it is again necessary to choose the test functions in an
adequate way. We use the spaces $W^{\sigma,b}_{1,1/l}\subset
S_{1,l}$, where $\sigma$ and $b$ can be taken arbitrarily large.
This choice cannot be illustrated by the example of a free scalar
field because its truncated $n$-point functions vanish
identically, except for $n=2$, and the two-point function, being
translation invariant, is unchanged by the  deformation. But as a
hint, we note that to test, e.g., the behavior of
$\varDelta_+\otimes_\theta\varDelta_+$ in the spacelike
directions, it is natural to use  test functions decreasing like
$\exp\{-|x|/l\}$, with $l\ll 1/\mu$, and whose Fourier transforms
behave at infinity no worse, because the twisting of the tensor
product intermixes the coordinate-space asymptotic behavior with
that in momentum space, as shows~\eqref{3*} and the explanatory
example given in Sec.~\ref{S4}.

\begin{theorem}\label{T3}.  If the assumption~\eqref{5.4} on the
existence of a mass gap holds, then for each test function  $f\in
W^{\sigma,b}_{1,1/l}$, the function
$w^{T,\theta}_{\,\bi_1\dots\bi_n}*f$ satisfies the inequalities
\begin{equation}
\left.|\partial^\kappa
(w^{T,\theta}_{\,\bi_1\dots\bi_n}*f)(x)|\right|_{x_1^0=\dots=
x_n^0}\le C_{\kappa,\bar l}\|f\|_{1/\bar l, b}
\exp\left\{-\frac{\mu R}{2(n-1)(1+3\mu \bar l)}\right\},
 \label{5.9}
\end{equation}
where  $\bar l>l$ and can be chosen arbitrarily close to $l$.
\end{theorem}

{\it Proof.} By  definition~\eqref{5.5}, we have
\begin{equation}
 w^{T,\theta}_{\,\bi_1\dots\bi_n}*f=
 w^T_{\,\bi_1\dots\bi_n}*h,\qquad\mbox{where}\quad
  \hat h(p)= \hat f(p)\prod_{j<k} e^{-\tfrac{i}{2}p_j\theta
p_k}.
 \label{5.10}
\end{equation}
The Fourier transformation  maps $W^{\sigma,b}_{1,1/l}$ onto the
space of functions analytic in the complex $(1/l)$-neighborhood of
the real space and satisfying, for each $\bar l>l$, the conditions
\begin{equation}
\sup_{|s|\le 1/\bar l}|g(p+is)|\le C_{\bar l,\bar b'}\prod_j e^{-
\bar b'|p_j|^{\sigma'}},\quad\text{where $|s|=\max_j|s_j|$}.
\notag
\end{equation}
Because $\sigma'>1$, Lemma~\ref{L2} obviously extends to this
space, and the function $\prod_{j<k} e^{-\tfrac{i}{2}p_j\theta
p_k}$ is hence its multiplier. Therefore, $h$ belongs to
$W^{\sigma,b}_{1,1/l}$ and depends continuously on  $f$. It
follows from the Cauchy theorem that the norms of $S_{1,l}$ and
$W^{\sigma,b}_{1,1/l}$ are related by $\|h\|_{N,\bar l}\le
C_{N,\bar l}\|h\|_{1/\bar l,\bar b}\,$. We conclude that if $f\in
W^{\sigma,b}_{1,1/l}$, then~\eqref{5.7} implies~\eqref{5.9},  and
Theorem~\ref{T3} is thus proved.

\begin{remark}\label{R5}. {\rm For simplicity, we have considered the
vacuum expectation values of products of deformed fields
$\prod_j\phi^\theta_{\bi_j}(x_j)$ with a common deformation
parameter $\theta$. However, an analogous  theorem holds for
products $\prod_j\phi^{\theta_j}_{\bi_j}(x_j)$ with  different
$\theta_j$. The proof is the same, but with a multiplier of a
different form than $\prod_{j<k} e^{-\tfrac{i}{2}p_j\theta p_k}$
in~\eqref{5.10}. As it is clear from the foregoing, the most
interesting case is that when $\theta_j$ differ only in sign.}
\end{remark}

\section{Conclusion}\label{S6}

The noncommutative deformation~\eqref{2.5} gives an interesting
example of quantum fields defined as operator-valued tempered
distributions on the Schwartz space $S$ and satisfying the
asymptotic commutativity condition previously proposed  for highly
singular nonlocal fields with analytic test functions in
$S^\beta=W^{1/(1-\beta)}$, where $\beta<1$.  It should be
emphasized that the asymptotic commutativity
principle~\cite{S99,FS} is not fully implemented  in the simplest
deformation of Wightman field theory considered here, because the
commutator   $[\phi^\theta_\bi(x_1),\phi^\theta_{\bi'}(x_2)]$ of
fields with equal deformation parameters  does not satisfy it.
This commutator decreases in the spacelike region in the same
fashion as  the Wightman functions, i.e., exponentially with the
damping factor depending on the threshold mass $\mu$. A more
sophisticated way of deformation is apparently required for the
deformed field theory to  meet fully the condition of asymptotic
commutativity and thus allow a consistent physical interpretation
as  nonlocal field theory.

Theorem~\ref{T2}  proved above can be supplemented by an
additional statement. The noncommutative deformation~\eqref{2.5}
enters an elementary length $\ell\sim \sqrt{|\theta|}$  into the
theory, and this length can be included in the characterization of
the behavior of the matrix elements $u_{\Psi,\Phi}$ of the field
commutator $[\phi^\theta_\bi(x_1),\phi^{-\theta}_{\bi'}(x_2)]$.
Namely, it can be shown that if the function $g$ in the
definition~\eqref{2.8} of the vector $\Phi$ belongs to
$S^{\beta}(\oR^{4n})$, where $\beta<1/2$, then the distribution
$u_{\Psi,\Phi}$ has a continuous extension to the space
$W^{2,b}(\oV)$, where $b=1/(2\ell^2)$. As proved in~\cite{S08},
such a property is also characteristic of the matrix elements of
the commutator $[\phi(x_1),\phi(x_2)]$, where $\phi(x)$ is the
deformed normal ordered square $:\varphi\star_\theta\varphi:(x)$
of a free scalar field $\varphi$.

Theorem~\ref{T3} shows, in particular, that incoming and outgoing
$n$-particle scattering states can be defined for the deformed
interacting fields in four-dimensional spacetime  in the usual
way~\cite{BLOT,J} without appealing to the wedge locality. In the
case of lower dimensions it should be combined with   Hepp's
idea~\cite{H} of using the so-called non-overlapping scattering
states. In fact, to prove the existence of the $\theta$-dependent
asymptotic states, it suffices to use a weaker version of
Theorem~\ref{T3} which employs  test functions with compact
support in momentum space and shows a decrease faster than any
power of $1/R$, but the strong version~\eqref{5.9} is essential to
understanding the analytic properties of the corresponding
S-matrix. The construction of the scattering matrix is a more
subtle and complicated problem which will be discussed in detail
in a subsequent paper. A preliminary analysis shows that the
arguments used for this purpose in~\cite{GL1,GL2,L11} can be
adapted to asymptotic commutativity. Then it makes sense to
consider a wider class of deformations, not necessarily preserving
the wedge locality. In this connection, it is worth noting that an
analogous deformation can be performed on nonlocal quantum fields
defined initially on $S^\beta$, where $\beta<1$, and subject to
the asymptotic commutativity  condition. Analogues of
Theorems~\ref{T2} and \ref{T3} can be shown to hold in this case
too, and this implies the existence of asymptotic scattering
states.

\section*{Acknowledgments} This paper was supported in part by
the the Russian Foundation for Basic Research (Grant No.~12-01-00865).

\end{document}